\documentclass[namedreferences]{SolarPhysics}
\usepackage[optionalrh,solaenum,natbib]{spr-sola-addons} % For Solar Physics 
\usepackage{graphicx}                    % For eps figures, newer & more powerfull
\usepackage{color}                       % For color text: \color command
\usepackage{url}                         % For breaking URLs easily trough lines
\usepackage[pdfborder={0 0 0 },urlcolor=blue,breaklinks]{hyperref} 
\ifx \doiurl  \undefined \def \doiurl#1{\href{http://dx.doi.org/#1}{\url{#1}}}\fi
\ifx \adsurl  \undefined \def \adsurl#1{\href{http://adsabs.harvard.edu/abs/#1}{\url{#1}}}\fi

\begin{document}

\begin{article}

\begin{opening}

\title{Helium Emissions Observed in Ground-Based Spectra of Solar Prominences}

%%%%%%%%%%%%%%%%%%%%%%%%%%%%%%%%%%%%%%%%%%%%%%%%%%%
%% Authors Names
%
\author{R.~\surname{Ramelli}$^{1}$\sep
        G.~\surname{Stellmacher}$^{2}$ \sep
        E.~\surname{Wiehr}$^{3}$ \sep
        M.~\surname{Bianda}$^{1}$        
       }

%%%%%%%%%%%%%%%%%%%%%%%%%%%%%%%%%%%%%%%%%%%%%%%%%%%
%% Runningheads
% 
\runningauthor{R. Ramelli {\it et al.}}
\runningtitle{He II in Prominences}

%%%%%%%%%%%%%%%%%%%%%%%%%%%%%%%%%%%%%%%%%%%%%%%%%%%
%% Affiliations 
%
  \institute{$^{1}$ Istituto Ricerche Solari, Locarno, Switzerland,
               email: \href{mailto:ramelli@irsol.ch}{ramelli@irsol.ch} 
                  and \href{mailto:mbianda@irsol.ch}{mbianda@irsol.ch}\\ 
             $^{2}$ Institute d'Astrophysique, Paris, France,
               email: \href{mailto:stell@iap.fr}{stell@iap.fr} \\
             $^{3}$ Institut f\"ur Astrophysik, G\"ottingen, Germany,
               email: \href{mailto:ewiehr@astro.physik.uni-goettingen.de} 
                          {ewiehr@astro.physik.uni-goettingen.de}\\
             }

%%%%%%%%%%%%%%%%%%%%%%%%%%%%%%%%%%%%%%%%%%%%%%%%%%%
%%% Abstract 
\begin{abstract}

The only prominent line of singly ionized helium in the visible spectral 
range, He\,{\sc ii}\,4686\AA, is observed together with the 
He\,{\sc i}\,5015\AA{} singlet and the He\,{\sc i}\,4471\AA {} triplet 
line in solar prominences. The Na\,D$_2$ emission is used as a tracer 
for He\,{\sc ii} emissions which are sufficiently bright to exceed the 
noise level near $10^{-6}$ of the disk-center intensity. The so selected 
prominences are characterized by small non-thermal line broadening and 
almost absent velocity shifts, yielding narrow line profiles without wiggles. 
The reduced widths [$\Delta\lambda_D/\lambda$] of He\,{\sc ii}\,4686\AA{} are 
1.5 times broader than those of He\,{\sc i}\,4471\AA{} triplet and 1.65 times 
broader than those of He\,{\sc i}\,5015\AA{} singlet. This indicates that the 
He lines originate in a prominence--corona transition region with outwards 
increasing temperature.

\end{abstract}

%%%%%%%%%%%%%%%%%%%%%%%%%%%%%%%%%%%%%%%%%%%%%%%%%%%
%% Keywords
%
\keywords{Prominences, Quiescent, Helium ionization}

\end{opening}
%-------------------------------------------------

%%%%%%%%%%%%%%%%%%%%%%%%%%%%%%%%%%%%%%%%%%%%%%%%%%%
%% Sections
%
\section{Introduction}%\label{s:?} 

As the second most abundant element, helium plays an essential role in
astrophysics. Nevertheless, its spectrum is poorly understood. The faint
He\,{\sc ii}\,4685.7\AA{} line is of particular interest, since it is the only 
important He\,{\sc ii} line that can be observed with ground-based telescopes 
allowing higher spectral resolution than currently achieved for EUV He\,{\sc ii} 
lines from space: \citet{stellmacher03} find that the SUMER spectrograph 
(designed for broad coronal lines) does not resolve the narrow emissions from 
cool prominences, even after application of a maximum instrumental profile.

He\,{\sc ii}\,4685.7\AA{} has been observed in prominences during eclipses 
\citep{sotirovski65,poletto67}. Line-profile analyses, based on moderately 
resolved spectra from coronographs \citep{hirayama72,hirayama74} indicate 
that He\,{\sc ii}\,4685.7\AA{} originates from the same (cool) prominence 
regions as the usually observed Balmer, He\,{\sc i} and metallic lines.
\citet{zirin59} observe in ''flare-like loop'' ({\it i.e.} highly active and 
''hot'') prominences the width of He\,{\sc ii}\,4685.7\AA{} larger than that 
of He\,{\sc i}\,4471.5\,\AA, however, they neither discuss quiescent (''cool'') 
prominences nor the atomic fine-structure broadening of He\,{\sc ii} which 
may explain most of that excess. A detailed analysis of the He\,{\sc i} and 
He\,{\sc ii} lines in solar prominences requires high spectral resolution, 
high signal-to-noise ratio and careful absolute calibration, which is 
difficult to achieve \citep[see][]{illing75}, but is possible 
with modern CCD techniques.

The high ionization and excitation energy of 25 and 48 eV suggests that 
He\,{\sc ii}\,4685.7\AA{} may preferentially occur in hot prominences 
($T_{\rm kin}>8000$K) which are observed to have a high He-to-Balmer 
emission ratio and to be highly structured \citep{stellmacher94}. These, 
however, usually show important velocity shifts which disperse the line 
profiles. We therefore suppose that the chance to measure 
He\,{\sc ii}\,4685.7\AA{} increases {\it if the emitted photons are 
concentrated in wavelength, i.e. yield narrow line profiles free from 
spatial Doppler shifts}. Such emissions should occur in prominences with 
negligible macro-velocities and low non-thermal line broadening, which are 
known to be bright in H${\alpha}$ though with a low He-to-Balmer ratio 
\citep{engvold71,stellmacher95}. Such prominences show significant Na\,D 
and Mg\,b emission and saturated H${\alpha}$ (and even H${\beta}$) profiles 
but did so far not allow a He\,{\sc ii}\,4685.7\AA{} line profile analysis 
\citep{stellmacher05}. 

\section{Observations}

In August 2011 we observed prominence emissions with the Gregory-Coud\'e 
telescope at the Locarno Solar Observatory (IRSOL: \opencite{ramelli06}). The 
faint He\,{\sc ii}\,4685.7\AA{} emission is hardly visible in the raw spectra and 
can only be detected after careful subtraction of the superposed aureole spectrum.
We therefore used the Na\,D$_2$ emission as a visual tracer for prominence candidates 
with ''spectral photon concentration'' and thus sufficient He\,{\sc ii} radiance. 
The so selected emission regions elevated only few arc-seconds over the limb and 
occurred at low solar latitudes $\varphi<30^\circ$. We oriented the spectrograph 
slit perpendicular to the solar limb by means of an image rotator, and pointed the 
telescope so that the light of the solar disk did not pass the aperture in the 
prime focus ({\it cf.}, slit-jaw images in Figure\,1). This avoids an illumination 
of the secondary and the two folding flat mirrors of the telescope thus 
reducing the stray-light \citep[{\it {\it cf.,}} Figure\,2 in][]{stellmacher70} - 
in our case by almost a factor of two.

In order to obtain a sufficiently high signal-to-noise ratio for the faint 
He\,{\sc ii} line, we chose a slit width of two arc-seconds and superposed 
spectra of (typically) 10 seconds exposure, yielding total integration times 
between 50 and 300 seconds. We used an unmasked ''e2v\,CCD\,55-30'' sensor 
mounted on a ZIMPOL camera \citep{ramelli10} and operated the ZIMPOL system 
only in its intensity mode (Stokes\,{\it I}). The flat-field was deduced 
from attenuated disk-center spectra which also served for the absolute 
calibration of the emissions. We took these spectra under identical 
conditions (exposure, slit width) as those for the prominence (and 
aureole) except for a defined attenuation of the high disk-center light 
level using carefully calibrated neutral filters. This allows to express 
the CCD counts of the prominence emission in terms of CCD counts at 
disk-center which finally are converted into absolute radiance 
[erg~s$^{-1}$~cm$^{-2}$~sterad$^{-1}$] using the tables of \citet{labs70}.
  
%% Figure 
\begin{figure} 
\centerline{\includegraphics[width=\textwidth,clip=]{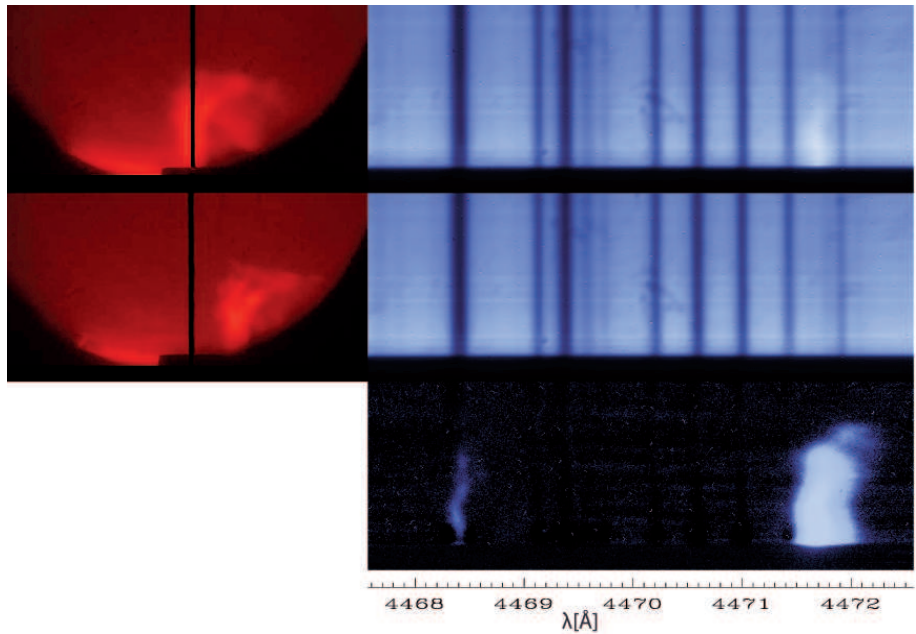}}
\caption{Raw spectra of a prominence (upper right panel) and the neighboring 
aureole (middle panel), i.e. the prominence displaced by a few arc-seconds from 
the slit (see left panels); difference image (lower panel) with prominence 
emissions, color scale adapted to the faint Ti\,II\,4468.4\AA{}, the strong 
He\,{\sc i}\,4471.5\AA{} emission thus over-saturated; the H${\alpha}$ slit 
jaw images show the chromosphere just at their bottoms, the solar disk is 
outside the field-stop in the primary focus, seen as round image boundary 
in the left panels.}
%\label{fig:1}
\end{figure}

The prominence emissions are superposed by an absorption spectrum from the 
solar disk (the ''aureole'') which is due to Rayleigh scattering mostly by 
dust particles on the telescope mirrors and less in Earth's atmosphere 
\citep[{\it cf.,} Figure\,2 in][]{stellmacher70}; it decreases in brightness 
with distance from the limb. We took such aureole spectra in the immediate 
vicinity of the respective prominence, normalized them to fit the spatial 
intensity distribution of the prominence spectra outside their emission 
lines, and subtracted them from the prominence spectral images (see 
Figure\,1). The ''zero level'' of the resulting emission spectra is 
disturbed by spurious remnants of the aureole lines arising from small 
(sub-pixel) spectral shifts between prominence and aureole exposure 
(spectrograph seeing). That ''$\lambda$-offset'' produces signatures 
that are anti-symmetrically shaped and can be described by the derivative of 
the aureole absorption profiles. We largely remove these signals by either 
adding or subtracting a fraction of the derivative of the respective aureole 
spectral profile. In the corrected spectra we finally select spatial prominence 
regions with strong emission and negligible Doppler structures.

On 2 and 4 August we observed exclusively Na\,D$_2$, He\,{\sc ii}\,4685.7\AA{} 
and the He\,{\sc i}\,4471.5\AA{} triplet line. After preliminary data reduction, 
we decided to include the He\,{\sc i}\,5015.7\AA{} singlet line as well. The 
first spectrum of that line (August\,9) showed the emission of the neighboring 
Fe\,{\sc ii}\,5018.4 line, drawing our attention also to the faint Ti\,{\sc ii}\,4468.4\AA{} 
line in the vicinity of He\,{\sc i}\,4471.5\AA{}. The occurrence of such ''chromospheric 
emissions'', usually observed only during solar eclipses, establishes our 
selection criterion and demonstrates the high quality of our observations. 
On August\,13 we added further metallic lines with significant emission in 
the table of \citet{sotirovski65}.

%% Figure 
\begin{figure} 
\centerline{\includegraphics[width=\textwidth,clip=]{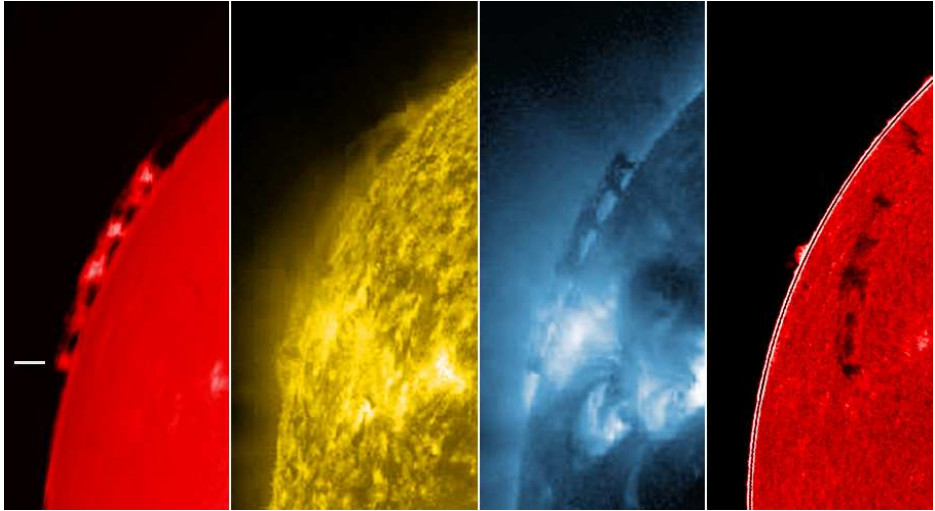}}
\caption{Images of the prominence chain at the east limb $+10^\circ 
<\psi<+35^\circ$ on August\,4, 2011, in H$\alpha$ from the Kanzelh\"ohe 
Observatory (left panel, the observed prominence at $14^\circ$\,N marked), in 
He\,{\sc ii}\,304\AA{} and Fe\,XVI\,335\AA{} from SDO/AIA (middle panels), together 
with the corresponding disk filament on August\,6 (right panel, Kanzelh\"ohe).}
%\label{fig:2}
\end{figure}

Among dozens of prominences occurring in the first half of August 2011, we found 
only few with markedly bright and narrow Na\,D$_2$ profiles, however, only four of 
these showed measurable He\,{\sc ii}\,4685.7\AA{} emission. The range of observed 
total  He\,{\sc i}\,4771.5\AA{} emission (radiance) $745<E(4471)<5120$ (see Table\,1), 
largely exceeds the values $E(4471)<320$ [erg s$^{-1}$ cm$^{-2}$ sterad$^{-1}$] by 
\citet{stellmacher97}. This shows that prominences with high He\,{\sc i} radiance 
also yield a sufficient He\,{\sc ii} emission above the noise level of few $10^{-6}$ 
of the disk-center brightness (see Figure\,3). Indeed, the chain of neighboring 
prominences occurring on August\,4 at the east limb between $10^\circ$N and $35^\circ$N 
contained two with directly visible Na\,D$_2$ emission, but only one ($14^\circ$N, 
marked in Fig.\,2) with Na\,D profiles narrow and bright enough to allow an observable 
He\,{\sc ii}\,4686 emission. Neither the H$\alpha$ or the He\,{\sc ii}\,304\AA{} 
appearance above the limb nor that of the corresponding filament on the disk and 
even not the cold prominence body, seen in the Fe\,XVI\,335\AA{} image as Lyman and 
He\,{\sc i} continuum absorption, give an indication for a peculiarity of that prominence 
(E,$14^\circ$N) favoring the detectability of a He\,{\sc ii}\,4685.7\AA{} emission. 

%% Figure 
\begin{figure}
\centerline{\includegraphics[width=0.5\textwidth,height=5cm,clip=]{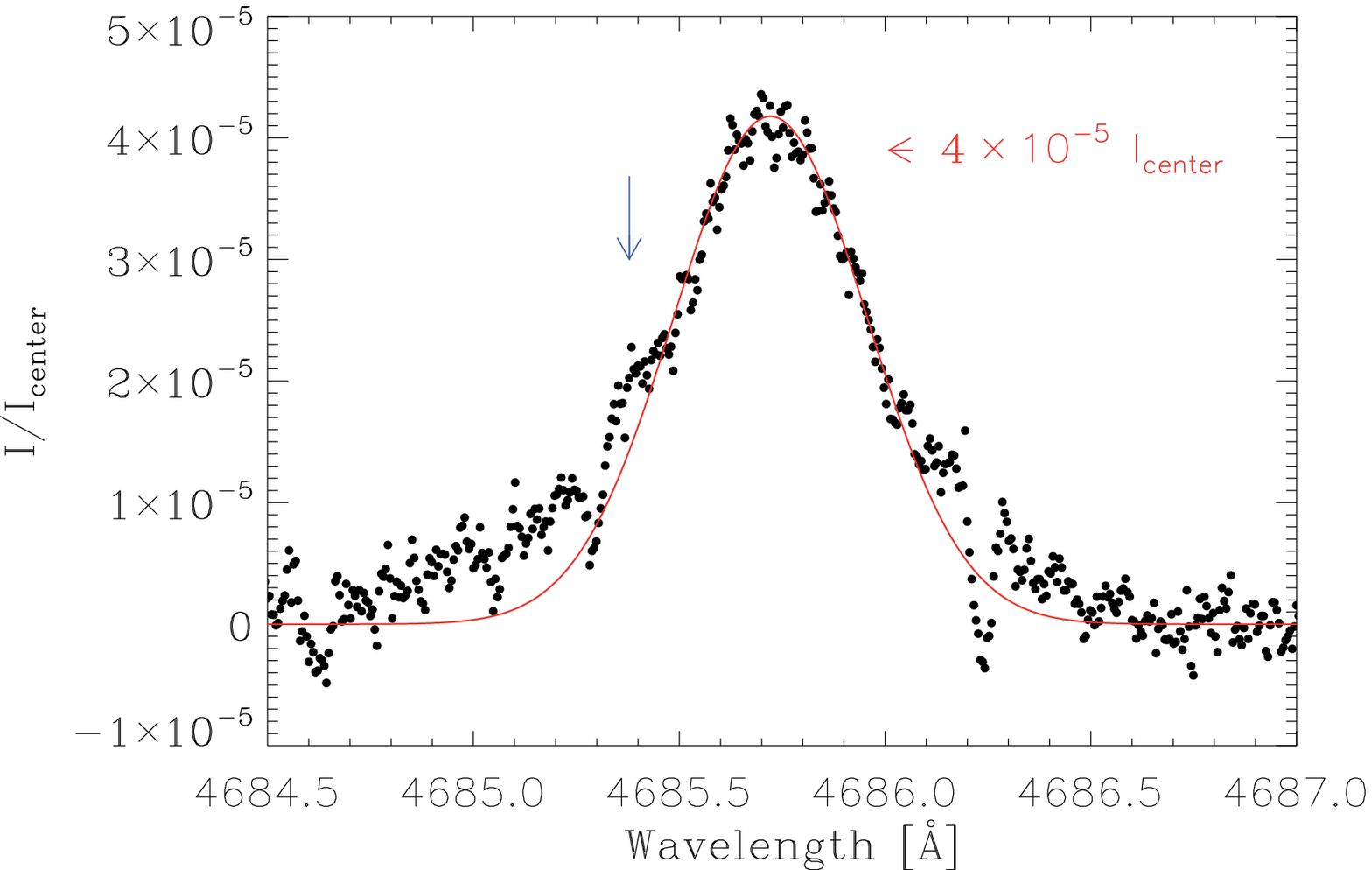}
\includegraphics[width=0.5\textwidth,height=5cm,clip=]{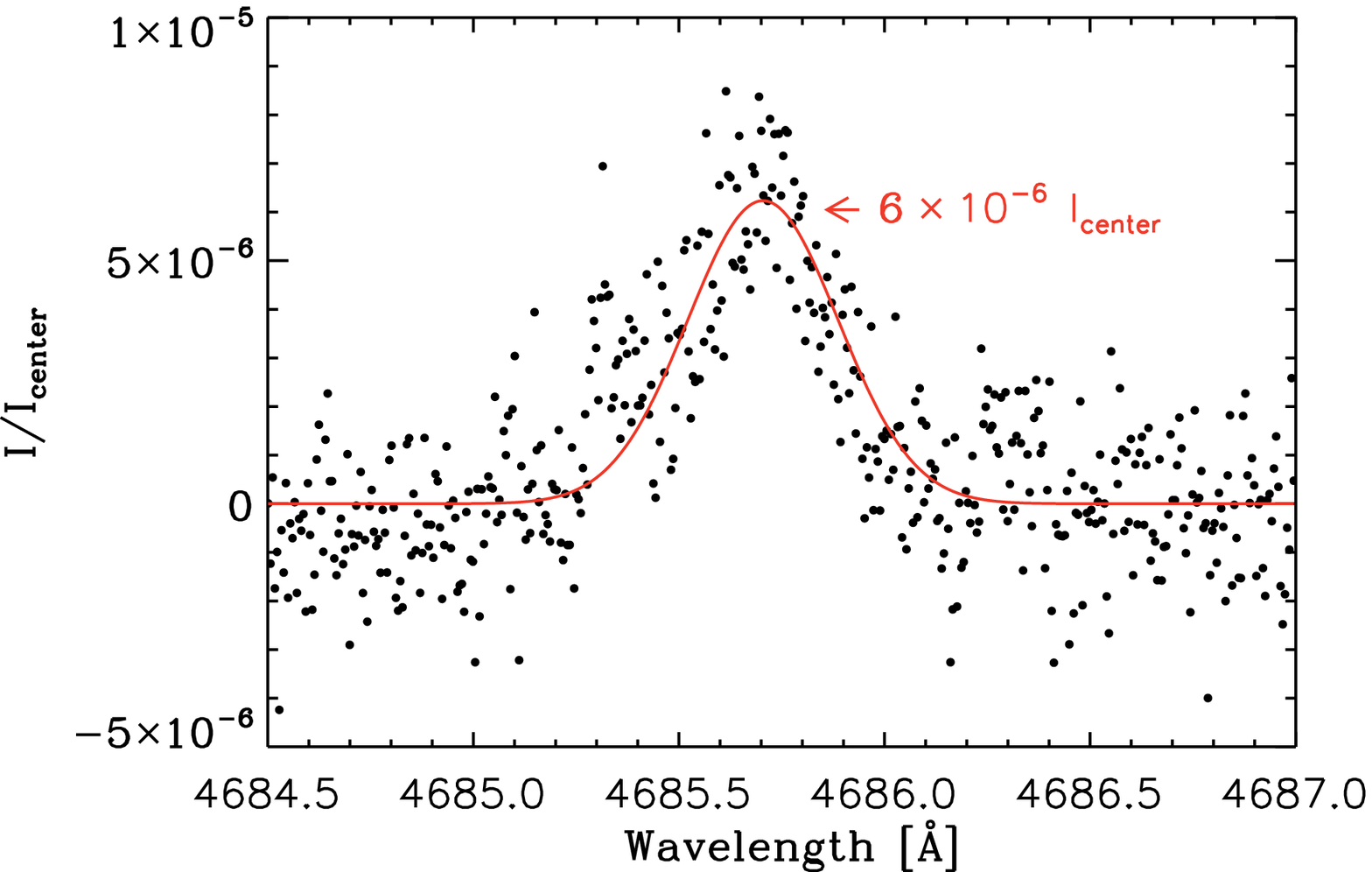}}
\centerline{\includegraphics[width=0.9\textwidth,clip=]{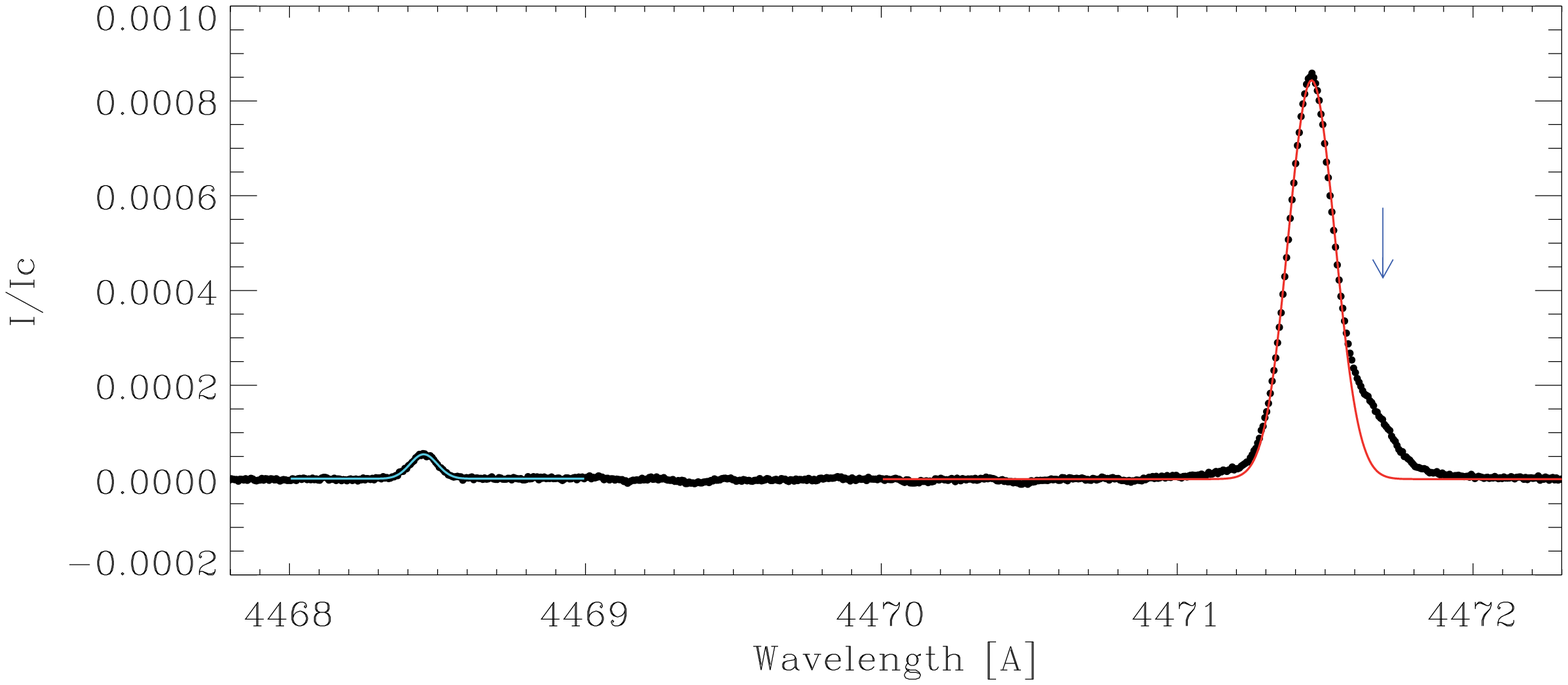}}
\caption{Strongest and faintest He\,{\sc ii}\,4685.7\AA{} profiles from 2 and 13 August
(note the different ordinate scales) together with Ti\,II\,4468.4\AA{} and 
He\I\,4471.5\AA{} ({\it cf.}, lower panel of Fig\,1). The colored lines give Gaussian 
fits to the upper profile parts. The atomic fine-structure components at 
4685.35 \AA{} and at 4471.7\AA{} (blue arrows) prove the high spectral 
resolution achieved.}
%\label{fig:3}
\end{figure}

\section{Results}
\subsection{Line Radiance}

Figure\,3 shows the strongest and the faintest He\,{\sc ii} emissions obtained at 
a noise level of $\pm1\cdot10^{-6}$ of the disk-center radiance. In Table\,1 we 
summarize the observed total line emission and compare them with observations 
by \citet{sotirovski65} and by \citet{poletto67}. Our low-latitude prominences 
($\varphi<30^\circ$) show much stronger radiance of the helium lines than the 
eclipse observations. The highest observed Na\,D$_2$ radiance of 1010 
[erg~s$^{-1}$~cm$^{-2}$~sterad$^{-1}$] (see Table\,1) corresponds to a total 
H${\alpha}$ emission of $2.5\times10^5$ [erg~s$^{-1}$~cm$^{-2}$~sterad$^{-1}$] 
which, in turn, is related to a large optical thickness of 
$\tau_0$(H$\alpha)\approx7.0$ \citep[according to observations 
by][]{stellmacher94,stellmacher05}; this proves that our criterion 
indeed selects thick prominences. 

We find mean radiance ratios $E(4471)/E(5015)=8.7$ for triplet-to-singlet and 
$E(4471)/E(4686)=51$ for triplet-to-He\,{\sc ii} ({\it cf.}, Table\,1). These ratio 
values are comparable to those obtained by \citet{sotirovski65} and by 
\citet{poletto67} from eclipse observations, although our prominences are much 
brighter. Hence, the radiance ratios seem to be valid over a large range of 
He emission.

%% Figure 
%
\begin{figure}
\centerline{\includegraphics[width=\textwidth,clip=]{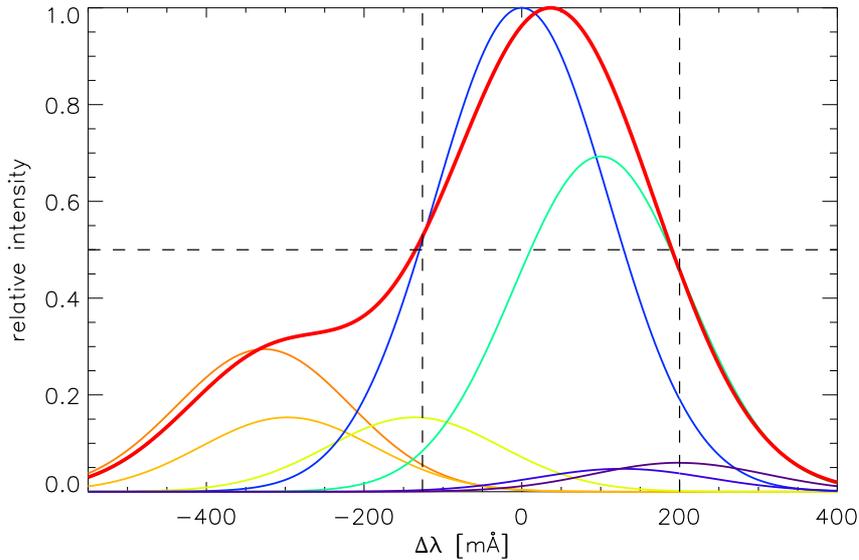}}
\caption{Superposition of the seven strongest (among 13) atomic fine-structure 
components of He\,{\sc ii}\,4685.7, each one with FWHM=260\,m\AA{} (chosen to fit
the August\,2 observation Figure\,3). The resulting convoluted profile (red line, 
max. normalized) with FWHM=325\,m\AA{} (dashed lines) is 1.25 times broader than 
the 'intrinsic' ({\it e.g.} blue) profile of each component.}
%\label{fig:4}
\end{figure}

\subsection{Line Widths}

The spectrograph slit of 0.25\,mm width (corresponding to 2~arc-seconds) 
yields a typical instrumental line broadening of 
$\Delta\lambda_e/\lambda\approx8\times10^{-6}$ which we apply to our observed 
line profiles. In addition, one has to consider the atomic fine-structure 
broadening: For He\,{\sc i}\,4471.5 five main components almost coincide 
($\Delta\lambda=19$m\AA); an additional faint component 210\,m\AA{} red-wards 
(well visible in the lower panel of Figure\,3) does not broaden the half-width, 
FWHM, of the composed profile. 

He\,{\sc ii}\,4685.7\AA{} is composed of 13 atomic fine-structure components 
leading to a significant line broadening. A corresponding deconvolution has to 
take into account that the various atomic fine-structure components have different 
intensity but a unique (''intrinsic'') width. He\,{\sc ii}\,4685.7\AA{} consists of 
two close main components ($\Delta\lambda=0.53$\,m\AA, blue line in Fig.\,4); 
a third one 100\,m\AA{} red-wards (green line in Figure\,4) broadens the FWHM 
of the composed profile (red line in Figure\,4) by a factor of 1.25 which is 
independent from the intrinsic profile width. Two atomic 
fine-structure components (orange and yellow lines in Figure\,4) produce 
a satellite at -350\,m\AA{} which is hardly visible in the upper left panel 
of Figure\,3 but does not influence the FWHM of the Gaussian fit to the 
upper part of the observed emission profile. 

%% Figure 
\begin{figure} 
\centerline{\includegraphics[width=\textwidth,clip=]{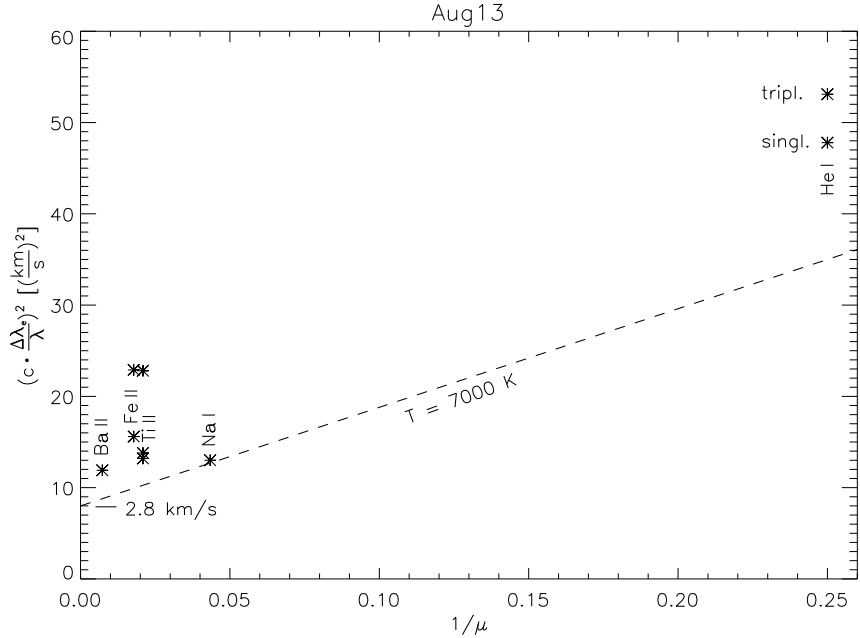}}
\caption{Velocity $v_o^2=(c\cdot \Delta\lambda_{\rm D}/\lambda)^2$ {\it versus} 
inverse atomic weight $1/\mu$ for the August\,13 prominence; the dashed line 
connects $v_o^2$(Na\,{\sc i}) with a reasonable $v_o^2$(H\,{\sc i}) corresponding 
to $[\Delta\lambda_e/\lambda]_H=3.7\times10^{-5}$ from \citet{stellmacher94}; 
the $v_o^2$ values of He\,{\sc ii} and of singly ionized metallic lines are 
located above that line; $v_o^2$(He\,{\sc ii})=173 is outside the ordinate range.}
%\label{fig:5} 
\end{figure}

Considering the He\,{\sc ii} fine-structure broadening, we find that {\it 
the He\,{\sc ii}\,4685.7\AA{} line is 1.5 times broader than the 
He\,{\sc i}\,4471.5\AA{} triplet line which, in turn, is 1.1 times broader 
than the He\,{\sc i}\,5015.7\AA{} singlet line} (Table\,2). 

\subsection{Kinetic Temperature}

The Doppler widths [$\Delta\lambda_{\rm D}$] from atoms of different weight generally allows one to
separate the thermal [$T_{\rm kin}$] from the (Maxwellian) non-thermal $(v_{\rm nth})$ line 
broadening: $v_o^2=(c\cdot\Delta\lambda_{\rm D}/\lambda)^2=2RT_{\rm kin}/\mu+v_{\rm nth}^2$, 
{\it if the lines are emitted in the same formation region}. Associating 
$[v_o^2(1/\mu)]_{\rm Na\,D}$ with $[v_o^2(1/\mu)]_{\rm He}$ from the smallest He\,{\sc i} 
(i.e. the singlet) line, we would obtain $T_{\rm kin}\approx10^4$K, a temperature too 
high for our selected dense, H${\alpha}$ bright and optically thick prominences. These 
typically show for hydrogen $[\Delta\lambda_{\rm D}/\lambda]_{\rm H}=3.7\cdot10^{-5}$ 
\citep{stellmacher94} which would yield $T_{\rm kin}=7000$\,K with the $v_o$ value 
observed for Na\,D$_2$ on August\,13 (dashed line in Figure\,5). The corresponding 
ordinate offset gives a small value $v_{\rm nth}=2.8$\,km/s confirming our criterion 
of 'spectral photon concentration' (see introduction). 

The $v_0^2$ values of the singly ionized metallic lines are found above the dashed 
line in Figure\,5 indicating an excess broadening. This result disagrees with 
\citet{landman85} who found Fe\,{\sc ii}\,5169 smaller than Mg\,{\sc i}\,b2 and b4. 
Also our observed $v_0^2$ values for He\,{\sc i} singlet, triplet and He\,{\sc ii} 
are above the dashed line [$T_{\rm kin}=7000$\,K; $v_{\rm nth}=2.8$\,km/s] in Figure\,5. 
Keeping $v_{\rm nth}=2.8$\,km/s, the widths of these three helium lines observed on 
August\,13 would correspond to $T_{\rm kin}$ values of 9400\,K, 10800\,K and 40000\,K, 
respectively; for constant $T_{\rm kin}=7000$\,K they would yield $v_{\rm nth}$ values 
of 7.3, 8.1, and 15\,km/s.

\section{Discussion}
\subsection{Line Radiance}

In order to correctly describe the helium spectrum in the prominence plasma, one 
has to consider the statistical equilibrium of ortho-, para- and ionized helium and 
their interaction ({\it e.g.} \opencite{labrosse10}). Observations of spectrally well
resolved line profiles are required for the understanding of the helium spectrum. 
The optically thin He\,{\sc ii}\,4685.7\AA{} line is poorly documented from observations. 
Its excitation is sensitive to radiation and collisions \citep{yakovkin71,labrosse10}, 
but also the influence of turbulence and of flows may enhance the He\,{\sc ii} emission 
\citep{jordan97,patsourakos02}. A blending with Ni\,{\sc i}\,4686.22\AA{}, mentioned by 
\citet{worden73} in their study of chromospheric emissions, is not indicated
in our prominence observations.

%+++++++++++++++++++++++++++++++
%% Table(1)
\begin{table}
\caption{Total line emission (radiance) [erg~s$^{-1}$~cm$^{-2}$~sterad$^{-1}$] in 
comparison with eclipse data from \citet{sotirovski65} and from \citet{poletto67}.}
\begin{tabular}{lcccccccc}                       % l: left, c: center, r: right
\hline 
position & W$22^\circ$N & E$14^\circ$N & E$30^\circ$N & E$31^\circ$N & E$23^\circ$N & Soti- & Poletto \\
obs.date       &2\,Aug.&4\,Aug.&9\,Aug.\,a&9\,Aug.\,b&13\,Aug.& rovski& + Rigutti\\
\hline \\
 He\,{\sc ii}\,4685.7         & 105  &  35  &   38  &   --  &   13  &  4.0  &  5.5  \\
 He\,{\sc i}\,4471.5\,(tr)    & 5119 & 1440 &  1940 &  1300 &  745  &  105  &  80   \\
 He\,{\sc i}\,5015.7\,(si)    &  --  &  --  &  198  &   180 &   82  &  19   &  --   \\ 
 Fe\,{\sc ii}\,5018.4         &  --  &  --  &  225  &   340 &   37  &   --  &  --   \\
 Ti\,{\sc ii}\,4468.4         &  85  &  185 &   64  &   105 &   31  &   --  &  --   \\
 Na\,D$_2$\,5890              & 1010 &  540 &  715  &   970 &   70  &   --  &  --   \\ \\

 tripl/He\,{\sc ii}           &   49 &   41 &   50  &   --  &   63  &  26   &   15  \\
 tripl/singl                  &  --  &  --  &   9.8 &  7.2  &  9.1  &  5.7  &  --   \\
 tripl/Na\,D$_2$              &  3.0 &  2.7 &   2.7 &  1.3  & 10.6  &   --  &  --   \\ \\
 \hline
 
\end{tabular}
\end{table}
%+++++++++++++++++++++++++++++++

Our radiance ratio $7.2<E(4471)/E(5015)<9.8$ (Table\,1) is in good 
agreement with $E(4471)/E(5015)=9.5$ from models with $T=8000$\,K and 
$n_{\rm H}\approx 10^{10}$ cm$^{-3}$ by \citet{heasley74}. However, the 
radiance ratio $41<E(4471)/E(4686)<62$ found in our study as well as in 
the eclipse data \citep{sotirovski65, poletto67} is smaller than 
$E(4471)/E(4686)=6950$ obtained from those models. The latter ratio 
would predict for our faintest prominence (August\,13 with $E(4471)=745$) 
a He\,{\sc ii}\,4685.7\AA{} radiance of $E(4686)\approx0.1$ 
[erg~s$^{-1}$~cm$^{-2}$~sterad$^{-1}$] far below the noise level - 
even for eclipse observations. Those models are isothermal; recent models, 
however, with prominence--corona transition region, PCTR, show an increased 
He ionization (\cite{labrosse04}). Our observed $E(4471)/E(4686)$ ratio is 
then in favor of an origin of He\,{\sc ii}\,4685.7\AA{} in the PCTR.

\subsection{Line Widths}

A further hint of the PCTR contribution to the He lines is given by the different 
line widths observed for the three atomic states: singly ionized, triplet, singlet. 
The reduced width [$RW=\Delta\lambda_{\rm D}/\lambda$[ of the He\,{\sc ii}\,4685.7\AA{} 
line (after deconvolution of the atomic fine-structure) is significantly larger than 
that of the He\,{\sc i} triplet line: $1.18<RW(4686)/RW(4471)<1.8$, and the triplet 
is broader than the singlet line: $1.07<RW(4471)/RW(5015)<1.09$ (see Table\,2). 
The latter result disagrees with observations by \citet{heasley75}, bearing in 
mind their lower spectral resolution. Since the He\,{\sc i} lines are even visible 
in our raw spectra (see Fig.\,1; in contrast to the faint He\,{\sc ii}\,4685.7\AA), 
the observed excess width $RW(4471)>RW(5015)$ is outside the error bars of at most 
$2\%$ (see Table\,2).

%+++++++++++++++++++++++++++++++
%% Table(2)
\begin{table}
\caption{Reduced Doppler widths $\Delta\lambda_{\rm D}/\lambda$~[10$^{-5}$]; the values 
for He\,{\sc ii}\,4685.7\AA{} after deconvolution of the atomic fine-structure; numbers 
in parentheses give uncertainties originating from the fitting procedure.}
\begin{tabular}{lccccc}                       % l: left, c: center, r: right
\hline                  
emission line                 &   2\,Aug.    &   4\,Aug.    &  9\,Aug.\,a  & 9\,Aug.\,b  &   13\,Aug.   \\
\hline \\
 He\,{\sc ii}\,4685.7         & 5.64\,(0.12) & 3.72\,(0.13) & 4.21\,(0.22) &      --     & 4.39\,(0.39) \\
 He\,{\sc i}\,4471.5\,(tripl) & 4.07\,(0.01) & 3.14\,(0.02) & 2.45\,(0.01) & 2.21\,(0.01)& 2.43\,(0.01) \\
 He\,{\sc i}\,5015.7\,(singl) &       --     &      --      & 2.24\,(0.03) & 2.07\,(0.04)& 2.28\,(0.02) \\
 Fe\,{\sc ii}\,5018.4         &       --     &      --      & 1.53\,(0.02) & 1.05\,(0.01)& 1.32\,(0.02) \\
 Ti\,{\sc ii}\,4468.4         & 3.28\,(0.04) & 2.30\,(0.04) & 1.17\,(0.05) & 0.92\,(0.02)& 1.21\,(0.08) \\
 Na\,D$_2$\,5890              & 3.17\,(0.60) & 2.13\,(0.03) & 1.36\,(0.02) & 1.08\,(0.01)& 1.20\,(0.01) \\ \\
 
He\,{\sc ii}/tripl            &     1.37     &     1.18     &     1.72     &     --      &     1.80    \\
tripl/singl                   &       --     &       --     &     1.09     &     1.07    &     1.07    \\
tripl/Na\,D$_2$           &     1.30     &     1.47     &     1.79     &     2.05    &     2.13    \\ \\

 \hline
\end{tabular}
\end{table}
%+++++++++++++++++++++++++++++++

The different widths of the He\,{\sc ii}, He\,{\sc i} triplet and singlet lines 
may be due to their formation in prominence regions of different temperature. 
Indeed, \citet{stellmacher03} observe ''hotter'' lines to be more pronounced 
in such prominence regions which show less radiance in ''cooler'' lines. 
Labrosse (private communication, 2011) finds among 100 models with a PCTR of 
$10^5$K ({\it cf.}, \opencite{labrosse04}) mean ratios of the reduced widths 
of $1.1<RW(4686)/RW(4471)<1.6$ and $1.03<RW(4471)/RW(5015)<1.2$, respectively, 
in good agreement with our results. The PCTR then seems to be essential for 
an explanation of the the observed ''hierarchy'' of He line widths.

The difference between the singlet and the triplet emission is explained 
by the fact that the lowest He\,{\sc i} energy level belongs to the singlet system 
which can thus be populated directly by EUV radiation, while the triplet levels 
are populated from the (singlet) ground state mainly by ionization and 
recombination. Collisional excitation plays a minor role in cool prominence 
regions, which are well visible as dark absorption features in the 
Fe\,XVI\,335\AA~image (see Figure\,2). These cool (and dense) prominence 
cores emit the narrow Na\,D lines which correspond to low T$_{\rm kin}$ and 
$v_{\rm nth}$ values as, e.g., the 7000\,K and 2.8\,km/s for August\,13 in Figure\,5.

%+++++++++++++++++++++++++++++++
\begin{table}
\caption{Reduced Doppler width $\Delta\lambda_D/\lambda$~[10$^{-5}$] and line 
radiance in absolute units [erg~s$^{-1}$~cm$^{-2}$~sterad$^{-1}$] of the metallic 
lines observed in the prominence at the east limb, $23^\circ$N on Aug.13, 2011; 
radiance data are compared with \citet{sotirovski65}.}
\begin{tabular}{cccccccccccc}                       % l: left, c: center, r: right
\hline 
ion            &Sr\,{\sc ii}&Ca\,{\sc i}& Ti\,{\sc ii}& Fe\,{\sc ii}&Ba\,{\sc ii}&Ti\,{\sc ii}&Mg\,{\sc i}I&Na\,{\sc i}& Na\,{\sc i}\\
$\lambda_0$\,[\AA] & 4215 & 4227 &  4468  &  4549  &  4554  &  4563  &  4571  & 5890  & 5896  \\
\hline \\ 
reduced\,width & 2.37 & 2.31 &  1.21  &  1.59  &  1.15  &  1.59  &  (0.87)&  1.20 & 1.14  \\ \\

obs. radiance   & 116  & 36.1 &  31.1  &  19.1  &  21.5  &  21.8  &   6.7  &   70  &  46   \\ 
Sotirovski     & 34.3 & 21.8 &  7.5   &   7.3  &   4.2  &   4.5  &   6.3  &  4.8  &  4.4  \\ \\
\hline 
\end{tabular}
\end{table}
%+++++++++++++++++++++++++++++++

The broad He lines, however, originate in hotter prominence regions where collisional 
excitation (and ionization) becomes effective. The observed ''hierarchy'' of excess 
broadening from He\,{\sc i} singlet to He\,{\sc i} triplet and He\,{\sc ii} lines may 
either be explained by differently hot threads or by a PCTR with outwards increasing 
temperature (or $v_{\rm nth}$). But the width excess of He\,{\sc ii}\,4685.7\AA{} 
corresponds to 40000\,K ({\it cf.,} end of Section\,3) which agrees with the formation 
temperature for He\,{\sc ii} lines calculated by \citet{gouttebroze09}. Hence, a 
temperature increase through the PCTR, as assumed in the models by \citet{heinzel01}, 
may be more realistic than an increase of $v_{\rm nth}$. Also a combined outwards increase 
of both, T$_{\rm kin}$ and $v_{\rm nth}$, might fit our observations.

The singly ionized metallic lines show a similar width excess as He\,{\sc i}\,4471\AA{}: 
In Figure\,5 their $v_0$ exceed $\approx$1.2 times the values expected for 
[$T_{\rm kin}=7000$\,K; $v_{\rm nth}=2.8$\,km/s]. Hence, the singly ionized metallic lines 
may be emitted from similar layers as the He\,{\sc i} triplet line. The rather tight 
relation of radiance and widths of Na\,D, He\,{\sc i}\,4471\AA{} and He\,{\sc ii}\,4686\AA{} 
(see Tables\,1 and 2) favors a PCTR surrounding each individual thread rather than the 
prominence as a whole.

%%%%%%%%%%%%%%%%%%%%%%%%%%%%%%%%%%%%%%%%%%%%%%%%%%%%%%%%%%%%%%%%%%%%%%%%%%%
%% Acknowledgments
%
\begin{acks} 
N.\,Labrosse kindly put the model means at our disposal. We thank  an unknown referee 
for fruitful comments, J.\,Hirzberger (MPS) for his code of fine-structure superposition 
and U.\,Nolte (GWDG) for the figure layout. The research at IRSOL is financially supported 
by the Canton Ticino, the foundation Aldo e Cele Dacc\`o, the city of Locarno, the local 
municipalities, and the SNF grant 200020-127329. R.\,R. acknowledges the foundation Carlo 
e Albina Cavargna for financial support.
\end{acks}

\newpage

%%% %%%%%%%%%%%%%%%%%%%%%%%%%%%%%%%%%%%%%%%%%%%%%%%%%%%%%%%%%%%
%% Bibliography
%
% Using BibTeX
%
\bibliographystyle{spr-mp-sola}
% %\bibliographystyle{spr-mp-sola-cnd} %% Alternative style: no title, no concluding page
\bibliography{helium}  
\end{article} 
\end{document}